\def\NAMEDBUILD{}
\def\LONGBUILD{}
\providecommand{\NEURIPSOPT}{preprint}
\documentclass{article}
\providecommand{\NEURIPSOPT}{dblblindworkshop}
\PassOptionsToPackage{hyphens}{url}
\usepackage[\NEURIPSOPT]{neurips_2026}
\ifdefined\NAMEDBUILD\else\workshoptitle{New In ML}\fi

\usepackage[utf8]{inputenc}
\usepackage[T1]{fontenc}
\usepackage{hyperref}
\usepackage{url}
\usepackage{booktabs}
\usepackage{amsfonts}
\usepackage{amsmath}
\usepackage{microtype}
\usepackage{graphicx}
\usepackage{natbib}

\newcommand{\qsixName}{Qwen3-0.6B}
\newcommand{\qfourName}{Qwen3-4B}
\newcommand{\dModel}{1024}
\newcommand{\nLayers}{28}

\newcommand{\ffBytes}{19}
\newcommand{\kvBytes}{4.1}
\newcommand{\crossover}{4{,}608}

\newcommand{\cacheKB}{112}
\newcommand{\cacheMega}{107}
\newcommand{\cacheSideW}{8.4}

\newcommand{\poolBytes}{23.1}
\newcommand{\crossoverPool}{3{,}584}
\newcommand{\holdPoolLong}{4}

\newcommand{\holdPoolMega}{1}
\newcommand{\holdPoolMegaComp}{99}
\newcommand{\stageOut}{2}
\newcommand{\stageBack}{6}
\newcommand{\tokenWire}{224}
\newcommand{\poolShare}{13}
\newcommand{\pairShare}{27}
\newcommand{\breakEven}{3}
\newcommand{\pairGemv}{2.3}
\newcommand{\pairBatched}{0.29}

\newcommand{\ffBytesBig}{149}

\newcommand{\crossoverBig}{36{,}480}

\newcommand{\cacheKBBig}{144}
\newcommand{\cacheMegaBig}{137}
\newcommand{\cacheSideWBig}{41.9}

\newcommand{\poolBytesBig}{159.9}
\newcommand{\crossoverPoolBig}{28{,}800}
\newcommand{\holdPoolLongBig}{23}

\newcommand{\holdPoolMegaBig}{4}

\newcommand{\stageOutBig}{5}
\newcommand{\stageBackBig}{9}
\newcommand{\tokenWireBig}{504}
\newcommand{\poolShareBig}{10}

\newcommand{\breakEvenBig}{7}
\newcommand{\pairGemvBig}{15.1}
\newcommand{\pairBatchedBig}{1.89}

\newcommand{\keyLateRel}{$3.6{\times}10^{-3}$}
\newcommand{\afdRequests}{12}
\newcommand{\afdLayers}{3}
\newcommand{\afdMinBatch}{2}
\newcommand{\afdLastQF}{11}
\newcommand{\afdLastComposed}{14}
\newcommand{\afdStranded}{1}
\newcommand{\backendGap}{$1.6{\times}10^{-2}$}

\newcommand{\splitRel}{$4.2{\times}10^{-3}$}

\newcommand{\shortCacheOver}{93 of 448}
\newcommand{\shortCacheNats}{9.7}

\newcommand{\combineOverflow}{$89$}
\newcommand{\contextLen}{one million}
\newcommand{\contextFraction}{65{,}536}
\newcommand{\shardPositions}{400}
\newcommand{\shardSpread}{30}
\newcommand{\shardZ}{$8.2{\times}10^{36}$}
\newcommand{\shardW}{$5.8{\times}10^{37}$}
\newcommand{\shardLogZ}{$85$}
\newcommand{\shardTrials}{200}
\newcommand{\shardErr}{$1.4{\times}10^{-7}$}

\newcommand{\srQ}{47.6}
\newcommand{\srK}{44.1}
\newcommand{\srV}{145.5}
\newcommand{\srQrand}{350.7}
\newcommand{\srKrand}{258.9}
\newcommand{\srVrand}{257.6}
\newcommand{\srQours}{43.3}
\newcommand{\srKours}{29.5}
\newcommand{\srVours}{55.9}
\newcommand{\kqOurs}{4.48}
\newcommand{\vqOurs}{3.33}

\newcommand{\kqRatio}{2.22}
\newcommand{\vqRatio}{1.15}

\newcommand{\truncRank}{48}
\newcommand{\truncCost}{+1.92}
\newcommand{\truncCostV}{+4.62}
\newcommand{\truncHalf}{+0.15}
\newcommand{\energyRank}{918}

\newcommand{\valPLzero}{1.2335}
\newcommand{\valPLone}{1.2400}
\newcommand{\valPLtwo}{1.2323}
\newcommand{\valPLmean}{1.2353}
\newcommand{\valSWzero}{1.2388}
\newcommand{\valSWone}{1.2381}
\newcommand{\valSWtwo}{1.2322}
\newcommand{\valSWmean}{1.2364}
\newcommand{\valKEzero}{1.2352}
\newcommand{\valKEone}{1.2365}
\newcommand{\valKEtwo}{1.2361}
\newcommand{\valKEmean}{1.2359}
\newcommand{\valQBzero}{1.2369}
\newcommand{\valQBone}{1.2403}
\newcommand{\valQBtwo}{1.2302}
\newcommand{\valQBmean}{1.2358}
\newcommand{\valQSzero}{1.2369}
\newcommand{\valQSone}{1.2377}
\newcommand{\valQStwo}{1.2332}
\newcommand{\valQSmean}{1.2359}

\newcommand{\splitMoveA}{0.0020}
\newcommand{\splitMoveMax}{0.0094}
\newcommand{\splitArms}{18}

\newcommand{\leakEffect}{-0.0071}
\newcommand{\leakSpread}{0.0014}
\newcommand{\leakRuns}{24}
\newcommand{\dPL}{-0.0011}
\newcommand{\dKE}{-0.0004}
\newcommand{\dQB}{-0.0006}
\newcommand{\dQS}{-0.0004}
\newcommand{\dSHQ}{-0.0090}
\newcommand{\dSQ}{+0.1003}
\newcommand{\maxDevA}{0.0011}
\newcommand{\nArms}{8}
\newcommand{\nRuns}{24}
\newcommand{\nSeeds}{3}
\newcommand{\gpuHoursTotal}{51}
\newcommand{\rangeSwap}{56}
\newcommand{\rangeSingle}{91}

\newcommand{\gapSW}{-0.0623}
\newcommand{\gapKE}{-0.0627}

\newcommand{\seedSpread}{0.0115}
\newcommand{\pairedSeedSpread}{0.0149}

\newcommand{\pairedSd}{0.0047}
\newcommand{\seedBand}{0.0117}
\newcommand{\spreadA}{0.0011}
\newcommand{\countA}{5}
\newcommand{\spreadB}{0.1093}

\newcommand{\queryEffect}{-0.0006}
\newcommand{\keyEffect}{-0.0004}
\newcommand{\swapEffect}{-0.0623}
\newcommand{\bothEffect}{-0.0004}
\newcommand{\shiftShare}{7}
\newcommand{\shiftEffect}{-0.0090}
\newcommand{\singleQ}{+0.1003}
\newcommand{\singleQzero}{+0.0979}
\newcommand{\singleQone}{+0.0970}
\newcommand{\singleQtwo}{+0.1061}
\newcommand{\swapEarly}{-0.1036}
\newcommand{\swapPeak}{-0.1722}

\newcommand{\convD}{+0.0142}
\newcommand{\convSeeds}{2}
\newcommand{\convDSpread}{0.0003}
\newcommand{\convUntrained}{+0.0746}
\newcommand{\convParams}{56.6M}
\newcommand{\convFrac}{9.5}
\newcommand{\convSteps}{3{,}000}
\newcommand{\convRefBpb}{0.9022}
\newcommand{\convArmBpb}{0.9166}

\newcommand{\depthFreeOne}{+0.0658}
\newcommand{\depthDOne}{+0.0123}
\newcommand{\depthKeptOne}{81}
\newcommand{\depthFreeThree}{+0.1750}
\newcommand{\depthDThree}{+0.0281}
\newcommand{\depthKeptThree}{84}
\newcommand{\depthFreeFive}{+0.3511}
\newcommand{\depthDFive}{+0.0414}
\newcommand{\depthKeptFive}{88}
\newcommand{\depthFreeSeven}{+0.4821}
\newcommand{\depthDSeven}{+0.0556}
\newcommand{\depthKeptSeven}{88}
\newcommand{\depthLayers}{24}
\newcommand{\depthHeld}{1--3}

\newcommand{\hybridName}{Qwen3.8-27B}
\newcommand{\hybridLayers}{64}
\newcommand{\hybridFull}{16}
\newcommand{\hybridOne}{+0.0181}
\newcommand{\hybridSeven}{+0.2081}
\newcommand{\hybridBase}{0.6806}
\newcommand{\meanQCost}{+0.8080}
\newcommand{\meanQBpb}{1.8698}
\newcommand{\betweenLayer}{67}
\newcommand{\meanQCostBig}{+1.1664}
\newcommand{\meanQBpbBig}{2.0895}
\newcommand{\betweenLayerBig}{73}
\newcommand{\dispSpread}{0.642}
\newcommand{\dispNorm}{0.384}
\newcommand{\commonShare}{57}
\newcommand{\dispSpreadBig}{0.616}
\newcommand{\dispNormBig}{0.329}
\newcommand{\commonShareBig}{61}
\newcommand{\dispSpreadHybrid}{0.519}
\newcommand{\dispNormHybrid}{0.214}
\newcommand{\commonShareHybrid}{66}
\newcommand{\Tol}{0.19}
\newcommand{\TolBig}{0.20}
\newcommand{\TolMoe}{0.12}
\newcommand{\TolMla}{0.89}
\newcommand{\TolBoth}{0.99}
\newcommand{\posEarly}{1.7}
\newcommand{\posLate}{7.3}
\newcommand{\posEarlyLo}{2}
\newcommand{\posLateHi}{1024}

\newcommand{\convParDBig}{+0.0427}
\newcommand{\convParFracBig}{44.6}
\newcommand{\convParUntrainedBig}{+2.2290}

\newcommand{\convBothD}{+0.0114}
\newcommand{\convBothFrac}{19.0}
\newcommand{\convBothDBig}{+0.0073}
\newcommand{\convBothFracBig}{18.2}
\newcommand{\convDBig}{+0.0088}
\newcommand{\convUntrainedBig}{+0.0364}

\newcommand{\convArmBpbBig}{0.7428}
\newcommand{\convFracBig}{9.1}
\newcommand{\runFloor}{0.010}
\newcommand{\armParams}{596.05M}
\newcommand{\armTokens}{372M}
\newcommand{\armSteps}{11{,}355}
\newcommand{\seqLen}{1{,}024}
\newcommand{\tokPerStep}{32{,}768}
\newcommand{\armLR}{$4{\times}10^{-3}$}
\newcommand{\armHours}{2.0 to 2.7}

\title{Q-First: Most of Attention Needs Only the Query\\
       in Disaggregated LLM Decoding}

\ifdefined\NAMEDBUILD
  \author{%
    WenJie Fan\\
    Yotta Labs\\
    \texttt{fanwj@mail.ustc.edu.cn} \quad \texttt{fanwenjie@yottalabs.ai}\\
    \texttt{https://github.com/fan-wenjie/qfirst} \quad \texttt{arXiv:2608.15473}
  }
\else
  \author{Anonymous}
\fi

\begin{document}
\maketitle

\begin{abstract}
\textbf{The problem.} Disaggregated decoding puts the KV-cache sweep on memory-optimised
hardware and the projections and feed-forward on compute-optimised hardware, then inherits the
decoder block's own ordering: attention runs first and the feed-forward consumes its output, so
within one sequence each device waits on the other. The usual repair, more sequences in flight,
costs one resident KV cache each --- what separating the devices was meant to avoid.

\textbf{The fix.} At a context of $j$ positions, $j-1$ are already cached, so all but one of
attention's score-and-combine terms follow from the query alone. Run the feed-forward first and
that query exists while the compute side still has work; the two then run concurrently, the
current key and value following as a write nothing waits on.

\textbf{What it costs.} Nothing structural: the decode is stated as a protocol, runs on the
attention kernel a stock framework ships, and verifies end to end on a trained checkpoint to
\splitRel{} relative, with no new operator, no changed tensor shape, no new hardware, and no
change to parameter count or arithmetic. Trained \nArms{} ways at \nSeeds{} seeds, no read point
in the family the protocol needs differs from the one that moves nothing by more than \maxDevA{}
bits per byte --- an order of
magnitude inside the \seedBand{} those seeds resolve, while the same runs resolve a sub-layer
exchange \rangeSwap{} times as large. A checkpoint in service moves to this read point by
fine-tuning \convFrac{}\% of its weights, at a residual \convD{} at \qsixName{} and
\convDBig{} at \qfourName{}.

\textbf{Why this read point.} Such a stack is the ordinary stack with its layer boundaries drawn
half a sub-layer over --- an identity, verified constructively --- so the freedom is exactly
where each of attention's three arguments is read, and only the query's position buys the
overlap. It is bounded: projecting every layer's query from the network's input costs
\singleQ{}, refuting a pre-registered threshold at every seed.
\end{abstract}
\section{Introduction}
\label{sec:intro}

A decoder layer reads two kinds of tensor: static weights, reused for every token and
arithmetic-dense, and the KV cache, which is state rather than weight --- touched once per byte
and memory-bound. Serving systems separate them onto hardware suited to each
\citep{lamina2024,adrenaline2025} and inherit from the decoder block the ordering that makes one
device wait on the other within a single sequence. This paper asks what the cache-side device
actually needs in order to start, and what it costs to give it that: the first answer is exact,
the second measured.

\paragraph{Contributions.}
\begin{itemize}
\item A decode protocol that gives the cache-side device the query and nothing else, so it may
      sweep while the compute side still has its feed-forward to run. Exact, no custom operator,
      on the attention kernel a stock framework ships; verified end to end on a trained checkpoint
      to \splitRel{} relative (Section~\ref{sec:protocol}). What a device returns is a mergeable
      aggregate, so one cache may be split across any number of devices and merged in any order
      (Section~\ref{sec:shard}), and two ordinary accelerators can be split by which \emph{kind} of
      weight they multiply by (Section~\ref{sec:twogpu}).
\item The measurement that makes this deployable: \textbf{making the query exist early does not
      cost quality.} In a block whose feed-forward runs first, moving the query's read point moves
      the model by \queryEffect{}~bits per byte against the block that moves none, the key's by
      \keyEffect{}, both by \bothEffect{} --- each inside a floor of \runFloor{} and an order of
      magnitude below what \nSeeds{} seeds resolve, \seedBand{}. Re-drawing the boundaries half a
      sub-layer over moves it \shiftEffect{}, also inside the floor. Trained from scratch at
      \qsixName{} at identical parameter counts, forward arithmetic, data order and seeds
      (Section~\ref{sec:cost}). A trained checkpoint reaches the same read point by fine-tuning
      $W_q$ alone, \convFrac{}\% of the model, leaving at most \convD{} (Section~\ref{sec:convert}).
\item The minimality argument the method is named for. The device sweeps keys already cached and
      the current key is a write nothing waits on, so exactly one of the attention's three read
      points has to move. Every alternative moves more --- two read points, or the graph itself, as
      the parallel block \citep{wang2021gptj,chowdhery2023palm} does in taking an $L$-layer stack
      from $2L$ sub-layers of serial depth to $L$ (Section~\ref{sec:algebra}).
\end{itemize}

\paragraph{What this paper does not claim.} No speedup is measured. The decomposition is exact
and the arrangement it admits is described in full, but what it buys needs two devices and a
decode loop; on one card the stream serialises everything regardless of what depends on what.
The training results are at \nSeeds{} seeds and one scale at three per cent of compute-optimal,
where a lead measures a disturbance to training rather than a quality reached. The conversion
result is one corpus at one budget, and bounds the residual cost rather than measuring it.

\section{The device needs only the query}
\label{sec:protocol}

Throughout, \emph{the device} means the memory-optimised side that holds the KV cache, and
\emph{the accelerator} the compute-optimised side that holds every weight. At decode step $j$
in one layer:

Write $u_j$ for the block's input at that step, $N_1$ for the attention's normalisation, and
$q_j = W_q N_1(u_j)$, $k_j = W_k N_1(u_j)$, $v_j = W_v N_1(u_j)$ for the query, key and value it
projects. In the rewired block the query still reads $u_j$ while the key and value read the
post-feed-forward stream, $k_j = W_k N_1(z_j)$ and $v_j = W_v N_1(z_j)$ with
$z_j = u_j + M(N_2\,u_j)$ (Section~\ref{sec:algebra}); the protocol below is stated for either
wiring. $o_{<j}$ and $\mathrm{lse}_{<j}$ are the attention vector over the cached positions
and their log-sum-exp.

\begin{center}
\begin{tabular}{lll}
\toprule
& sends & computes \\
\midrule
accelerator & $q_j$                                        & the feed-forward, then $k_j$ and $v_j$ \\
device      & $o_{<j}$ and $\mathrm{lse}_{<j}$             & attention over the $j-1$ cached positions \\
accelerator & $k_j, v_j$ \emph{(a cache write, unblocking)} & the combine below \\
\bottomrule
\end{tabular}
\end{center}

The device attends over the cache alone and returns the attention vector among those positions
with their log-sum-exp. The accelerator, which computed $q_j$ and $k_j$ itself, finishes with the
standard combination of two partial softmax states,
\begin{equation*}
s_{jj} = \frac{q_j \cdot k_j}{\sqrt{d}},
\qquad
\mathrm{out} = \mathrm{lerp}\big(v_j,\; o_{<j},\; \sigma(\mathrm{lse}_{<j} - s_{jj})\big),
\qquad \mathrm{lerp}(x, y, t) = (1-t)\,x + t\,y .
\end{equation*}

\paragraph{Overflow, where it actually is.} Not in the combine: written as $1/(1+e^{-x})$ the
weight is correct at every magnitude in IEEE arithmetic, agreeing with a library sigmoid
bit-for-bit at two million points across $[-120,120]$ and finite out to $\pm 800$. The
exponential form it replaces returns NaN past \combineOverflow{} in single precision. Three
places do need care --- the running log partition, how a fused kernel's padded log-sum-exp is
indexed, and the sign of $s_{jj}-\mathrm{lse}_{<j}$, which is not one-directional on a trained
checkpoint. Each is stated with its failure in the released code.

\textbf{The device's input is the query alone}, which holds of any decoder: the sweep is over
positions already in the cache and needs neither the current key nor value. What differs between
blocks is \emph{when} the query exists. The current key and value arrive afterwards as a cache
write nothing waits on, so the sweep runs concurrently with the feed-forward producing $v_j$ and
what stays serial is a combination of two vectors. No custom operator is needed: the
memory-efficient attention of \citet{dao2022flashattention} returns the log-sum-exp exactly, and
we verify the step against fused attention to \splitRel{} relative on a trained checkpoint. Two
caveats: that backend has no grouped-query kernel \citep{ainslie2023gqa}, so keys and values
must be expanded first; and PyTorch's \emph{flash} entry point returned a second output matching
no hand-computed log-sum-exp under either layout.

\paragraph{The intervention is one send, moved earlier.} Nothing above asks for a new operator,
a new kernel, or a tensor of a different shape. The sweep is $O(j\,d)$ per head and needs the
query alone; what remains after it is a rank-one combination, $O(d)$, one $j$-th of the layer's
attention work at a context of $j$: one \contextFraction{}th at 65{,}536 positions, and
proportionally less at \contextLen{}. A deployment obtains the overlap by
computing $q_j$ from the block's input and putting it on the wire at that moment rather than
after the layer's other arithmetic; the key and value follow whenever they are ready.

What has to change to make that possible is the block's ordering, not its parts. In an ordinary
block the attention runs first, so its query is available immediately --- and that is exactly the
problem, because the feed-forward then waits on the attention's output and the two devices
alternate rather than overlap. Putting the feed-forward first, with the query still reading the
block's input, leaves every component, shape and parameter where it was and leaves both sides
with work to do at once; Section~\ref{sec:algebra} shows the re-bracketing is exact.

\section{A two-accelerator arrangement}
\label{sec:twogpu}

The protocol says what may be computed when, not where, and the arrangement it admits is worth
stating because a reader will otherwise assume near-memory hardware that does not yet exist.
Two ordinary accelerators suffice, split by \emph{what kind of weight they multiply by} rather
than by layer or head: the MLP side holds the static weights and runs every projection and
feed-forward, the attention side holds the KV cache and does nothing but sweep it. This is the
split model-attention disaggregation already builds \citep{lamina2024,zhu2025megascale,stepfun2025step3}; what changes is only
what the cache side is given and when.

One layer, one decode step $j$, reading down:

\begin{center}
\small
\begin{tabular}{@{}p{0.30\textwidth}p{0.38\textwidth}p{0.22\textwidth}@{}}
\toprule
MLP side & attention side & shared, both can reach \\
\midrule
$q_j \rightarrow$ & receives $q_j$ & \\
computes MLP, then $k_j, v_j$ & sweeps the cache for $o_{<j}$, $\mathrm{lse}_{<j}$ & \\
 & parks the pair; stages layer $L{+}1$'s KV from its own VRAM & $o_{<j}$, $\mathrm{lse}_{<j}$ \\
reads the pair, combines with $v_j$ & & \\
$k_j, v_j \rightarrow$ & appends to layer $L$'s cache & \\
\bottomrule
\end{tabular}
\end{center}

Three properties of this arrangement are not incidental.

\textbf{The result is parked, not pushed.} The attention side writes $o_{<j}$ and
$\mathrm{lse}_{<j}$ to memory the MLP side can read and moves on rather than waiting for a
handshake, so the two are coupled by data and not by arrival order.

\paragraph{Across layers, not only within one.} Read one layer at a time, the sweep runs beside
the feed-forward of the same layer; drawing the stack one half-layer over makes it span layers
instead, which is the form to hand an implementer:
\begin{equation*}
[\,\mathrm{Attn}_1\,]\;
[\,\mathrm{MLP}_1 \parallel \mathrm{Attn}_2\,]\;
[\,\mathrm{MLP}_2 \parallel \mathrm{Attn}_3\,]\;\cdots\;
[\,\mathrm{MLP}_{N-1} \parallel \mathrm{Attn}_N\,]\;
[\,\mathrm{MLP}_N\,]
\end{equation*}
Each middle stage holds one feed-forward and the \emph{next} layer's attention, both reading the
stage's input, so layer $i+1$'s query exists before layer $i$'s feed-forward runs. Only two of
the $N+1$ stages lose the concurrency: the first has no feed-forward before its attention, the
last no attention after its feed-forward. The re-bracketing is exact and verified bit-identical
(Section~\ref{sec:algebra}).

\paragraph{The compute side is a preemptible pool.} Sweep and feed-forward may run at once,
neither being downstream of the other, and the pool holds no request state so it can be taken
and released at will (Section~\ref{sec:master}). We do not claim either conceals the other:
which finishes first depends on the context.

At batch one both sides are bandwidth-bound, so the comparison is bytes per layer per step: the
feed-forward reads its weights whatever the context, the sweep reads \kvBytes{}~KB per cached
position. One row per model, every figure this section quotes:

\begin{center}
\small
\begin{tabular}{lrrrr}
\toprule
model & feed-forward & sweep holds the pool & crossover & pooled \\
\midrule
\qsixName{}  & \ffBytes{}~MB    & \holdPoolMega{}\% at \contextLen{}, \holdPoolLong{}\% at 128K
             & \crossover{}    & \crossoverPool{} \\
\qfourName{} & \ffBytesBig{}~MB & \holdPoolMegaBig{}\% and \holdPoolLongBig{}\%
             & \crossoverBig{} & \crossoverPoolBig{} \\
\bottomrule
\end{tabular}
\end{center}

Crossover is where the two are equal; pooling the key and value projections with the
feed-forward moves it to the last column. The remaining \holdPoolMegaComp{}\% is a device holding
weights it cannot put down, which against a preemptible pool is another request's time. None of
this is measured: it is a scheduling consequence of the contract in Section~\ref{sec:master}, and
what decides whether it pays is the mix of contexts a deployment sees.

\section{What each side holds}
\label{sec:master}

A deployment has to break the symmetry: one side owns the request. Which side is settled by
where the state is pinned and by what a shared pool may hold, not by which component is
stateless --- a database holds more state than the service in front of it and is still the
callee.

\begin{center}
\small
\begin{tabular}{llp{0.50\textwidth}}
\toprule
held by & what & why it cannot sit on the other side \\
\midrule
cache side & the KV cache & migration is not an option, so a request is pinned for its lifetime
to the device holding it \citep{kwon2023pagedattention}. \\
cache side & $W_q$ & a query computed in the pool would arrive with the reply, no earlier than
the feed-forward it was meant to overlap. \\
cache side & $W_o$ & $W_q$ applies to $u' = z + W_o a$, so $W_o$ in the pool puts a round trip in
front of \emph{every} sweep --- the ordering this arrangement exists to remove, one layer down. \\
\midrule
pool & the feed-forward & at batch one it reads its whole weight for one token's arithmetic, and
only batching across \emph{different} requests amortises that \citep{yu2022orca}. A pool is
called, and cannot own one request's control flow without giving up the batching it exists
for. \\
pool & $W_k, W_v$ & inlined into the same call so the pool returns cache entries rather than a
hidden state the caller must send back. Nothing writes into another device's memory. \\
\bottomrule
\end{tabular}
\end{center}

\begin{center}
\small
\begin{tabular}{lrrrrrr}
\toprule
& \multicolumn{2}{c}{KV cache} & \multicolumn{2}{c}{held per layer} & \multicolumn{2}{c}{on the wire} \\
\cmidrule(lr){2-3}\cmidrule(lr){4-5}\cmidrule(lr){6-7}
model & KB/token & GB at 1M & cache side & pool & KB/stage & KB/token \\
\midrule
\qsixName{}  & \cacheKB{}    & \cacheMega{}    & \cacheSideW{}    & \poolBytes{}    & \stageOut{}/\stageBack{}       & \tokenWire{} \\
\qfourName{} & \cacheKBBig{} & \cacheMegaBig{} & \cacheSideWBig{} & \poolBytesBig{} & \stageOutBig{}/\stageBackBig{} & \tokenWireBig{} \\
\bottomrule
\end{tabular}
\end{center}

\noindent One call per stage, with the state on the wire:
\begin{equation*}
\text{cache} \xrightarrow{\;u\;} \text{pool}
\xrightarrow{\;z = u + M(N_2 u);\; k, v \text{ from } N_1(z)\;} \text{cache}
\xrightarrow{\;\text{sweep with its own } q\;} u' = z + W_o a .
\end{equation*}
A pool holding no request state is preemptible: a stalled caller stops calling and nothing of the
request is inside to protect.

\section{The cache may be split anywhere}
\label{sec:shard}

The two values the device returns are not an ad hoc pair. For a set $S$ of cached positions
with scores $s_i$ and values $v_i$, write
\begin{equation*}
Z(S) = \sum_{i \in S} e^{s_i}, \qquad
W(S) = \sum_{i \in S} e^{s_i} v_i, \qquad
O(S) = W(S) / Z(S),
\end{equation*}
and $(W, Z)$ adds componentwise over disjoint sets: the online-softmax recurrence
\citep{milakov2018online} behind single-pass fused attention \citep{dao2022flashattention} and
sequence-parallel merging \citep{liu2023ring}. The contribution is the coordinates, not the
algebra. A device returns $(O, \log Z)$ --- the same pair normalised by
its total, the total kept in logs --- and the merge there is the combine of
Section~\ref{sec:protocol}:
\begin{align*}
O(A \cup B) &= \mathrm{lerp}\big(O(B),\, O(A),\, \sigma(\log Z(A) - \log Z(B))\big), \\
\log Z(A \cup B) &= \mathrm{logaddexp}\big(\log Z(A),\, \log Z(B)\big).
\end{align*}

A cache split across any number of devices, in any way, swept independently and merged in any
order therefore returns what one sweep over the whole would: \shardTrials{} random four-way
splits agree to \shardErr{}. No split has to be balanced, contiguous or known in advance, and no
device needs to know what another holds.

One asymmetry is worth an implementer's attention: the identity is degenerate. An empty shard is
$\log Z = -\infty$ and a vector never read, so a device with nothing to contribute may return
whatever is in its buffer, provided its log partition says so.

\section{Converting a trained checkpoint}
\label{sec:convert}

Everything above describes a decode a deployment would have to build a model for. Trained from
scratch the read point costs nothing this campaign can measure (Section~\ref{sec:cost}), but
that obliges a reader to retrain before any of it is usable. This section asks the question an
existing deployment actually has: what it costs to move a checkpoint already in service to the
early-query read point. Qwen3-0.6B is rewired by hooks --- the block's own arithmetic is
untouched, layer $\ell$'s query reads $h_{\ell-1}$ instead of $x_\ell$, its key and value do
not move, and layer $0$ is exempt because its attention reads the embedding under either
wiring. Every comparison in this paper is at that read point and that coverage --- one
half-layer back, every layer the wiring can reach --- except the depth sweep, which varies
it, and the two named boundary arms, which is what they are for. Only $W_q$ is then trained, \convParams{} of the model or \convFrac{}\%; everything
else, including the tied embedding and therefore the readout, stays frozen.

The reference must receive the same fine-tune. Scored against the checkpoint evaluated
zero-shot, the converted-and-repaired model \emph{beats} what it was converted from, because
the fine-tune adapts it to a corpus the checkpoint never saw and that adaptation is worth
several times the conversion. So the quantity is

\begin{equation*}
d = (\text{converted} + \text{SFT}) - (\text{unconverted} + \text{the same SFT}),
\end{equation*}
at identical repair set, schedule, data order and seed.

\begin{center}
\small
\begin{tabular}{lcccc}
\toprule
& trainable & untrained & after \convSteps{} steps & $d$ \\
\midrule
\qsixName{}  & \convFrac{}\%    & \convUntrained{}    & \convArmBpb{}    & \convD{} \\
\qfourName{} & \convFracBig{}\% & \convUntrainedBig{} & \convArmBpbBig{} & \convDBig{} \\
\bottomrule
\end{tabular}
\end{center}

Rewiring costs \convUntrained{}~bits per byte before any training at \qsixName{}; fine-tuning
$W_q$ alone removes most of it and leaves \convD{}, reproduced across \convSeeds{} seeds to
\convDSpread{}. The residue is expected: the converted query must reproduce
$W_q x = W_q h + W_q M(h)$ from $h$ alone, and the second term is not linear in $h$, so no
retrained linear $W_q$ is exact.

Two things this does not say. It is not a quality comparison between arrangements --- the
parallel block trains perfectly well from scratch, and nothing here measures that. And equal
steps is not equal difficulty: the converted arm spends part of its budget undoing the
conversion while the reference spends all of its adapting, so \convD{} is an upper bound on
the residual cost, not a measurement of it.

The recipe is therefore short: rewire the read point, retrain one projection per layer, leave
everything else including the readout frozen. Adding the output projection --- the other weight
the cache side already holds, so still no new parameter and no new inference-time matmul ---
takes $d$ to \convBothD{} at \qsixName{} and \convBothDBig{} at \qfourName{}, for
\convBothFrac{}\% and \convBothFracBig{}\% trainable. Larger repair sets, alternative wirings
for the query path, and the forward-only diagnostics bounding them were also run; none beats
this, and none is reported here.

\paragraph{How far back the query may be read.} The arrangement above overlaps one feed-forward,
and one half-layer opens it, so in a dense stack nothing deeper buys overlap. A pattern that
alternates several linear-attention layers with one softmax-attention layer offers a longer
shadow, and there deeper reads do open more of it --- but the residual after repair rises with
them, \depthDOne{} at one half-layer against \depthDSeven{} at seven, and the residual is what a
deployment carries. Half a layer is therefore the conservative choice under both block patterns,
and it is where this paper measures.

\begin{center}
\small
\begin{tabular}{ccccc}
\toprule
half-layers & of the shadow & untrained & $d$ & removed \\
\midrule
1 & 12.5\% & \depthFreeOne{}   & \depthDOne{}   & \depthKeptOne{}\%   \\
3 & 37.5\% & \depthFreeThree{} & \depthDThree{} & \depthKeptThree{}\% \\
5 & 62.5\% & \depthFreeFive{}  & \depthDFive{}  & \depthKeptFive{}\%  \\
7 & 87.5\% & \depthFreeSeven{} & \depthDSeven{} & \depthKeptSeven{}\% \\
\bottomrule
\end{tabular}
\end{center}
\noindent Untrained is the displacement before any repair, $d$ what survives it, and the last
column their ratio: the repair removes more at depth and still leaves more.

\section{What it costs to make the query exist early}
\label{sec:cost}

The protocol needs the query before the feed-forward; whether a block can be trained that way is
a separate question, answered here. Every comparison holds the computation graph, the components,
their shapes, the parameter count, the forward arithmetic, the data order and the seed fixed, and
moves the read point alone. The arms fall into two families by whether a block's feed-forward
consumes that block's own attention: in family A it does not, the feed-forward running first or
the two running from the same input; in family B it does, the ordinary arrangement. Every run is at
\nSeeds{} seeds, all printed, because a mean hides which seed the spread came from.

\begin{figure}[t]
\centering
\includegraphics[width=0.86\textwidth]{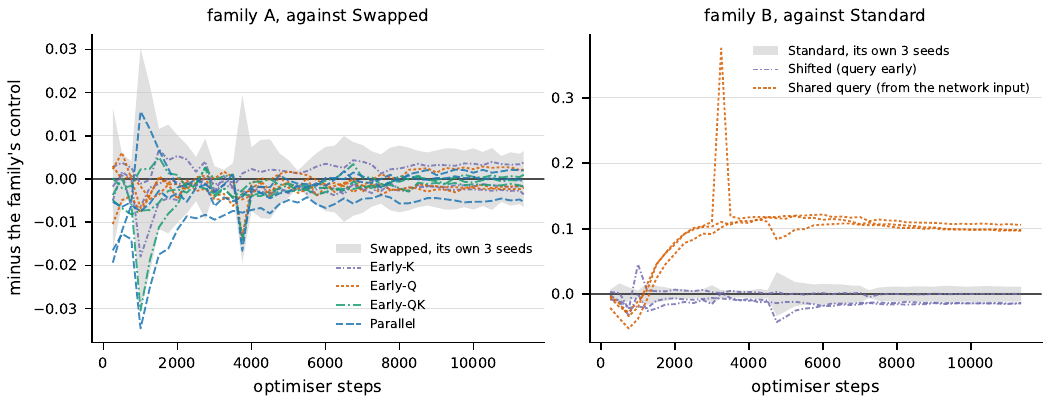}
\caption{Each family against its own control, every seed: family A against Swapped, family B
against Standard. Grey is the control's own spread across seeds, the width inside which nothing
is a difference; the panels do not share an axis, their ranges differing ninefold. Left: no read
point leaves the band by much or for long and all four are inside it over the last third. Right:
Shifted-Q sits just below the control at \shiftEffect{}, the shared query source far above it at
\singleQ{}.}
\label{fig:curves}
\end{figure}

\paragraph{Family A: the feed-forward does not consume this block's attention.} Exchange the two
sub-layers and the query and key may each read the residual stream before the feed-forward or
after it; the value always reads after, since reading it early too is the parallel arrangement.
The control is Swapped, the corner that moves no read point.
\begin{center}
\small
\begin{tabular}{llccccc}
\toprule
arm & reads early & seed 0 & seed 1 & seed 2 & mean & vs Swapped \\
\midrule
Swapped   & nothing                  & \valSWzero{} & \valSWone{} & \valSWtwo{} & \valSWmean{} & --- \\
Early-K   & the key                  & \valKEzero{} & \valKEone{} & \valKEtwo{} & \valKEmean{} & \dKE{} \\
Early-QK  & query and key            & \valQSzero{} & \valQSone{} & \valQStwo{} & \valQSmean{} & \dQS{} \\
Early-Q   & the query                & \valQBzero{} & \valQBone{} & \valQBtwo{} & \valQBmean{} & \dQB{} \\
Parallel  & \emph{a different graph} & \valPLzero{} & \valPLone{} & \valPLtwo{} & \valPLmean{} & \dPL{} \\
\bottomrule
\end{tabular}
\end{center}
The \countA{} arms span \spreadA{} against a floor of \runFloor{}, and none differs from the
control by more than \maxDevA{}, and the same holds throughout training
(Figure~\ref{fig:curves}). Nothing
here is distinguishable from anything else, which is the only claim the table makes and the one
the protocol needs: where the feed-forward does not wait on its attention, moving the read point
is a change this campaign cannot find.

\paragraph{Family B: it does.} The ordinary ordering, with two arms that change where the query
comes from without exchanging anything: the shifted stack re-draws the boundaries half a
sub-layer over as Section~\ref{sec:algebra} does, taking each query from the input of the
preceding feed-forward; the shared-query arm projects every layer's query from the network's
input.
\begin{center}
\small
\begin{tabular}{llc}
\toprule
arm & the query is read from & vs Standard \\
\midrule
Shifted-Q & the preceding feed-forward's input & \dSHQ{} \\
Single-Q  & the network's input                & \dSQ{} \\
\bottomrule
\end{tabular}
\end{center}
This family spans \spreadB{}, and the two treatments differ from the control in opposite
directions: the shifted stack by \shiftEffect{} and the shared source by \singleQ{}
(\singleQzero{}, \singleQone{} and \singleQtwo{} by seed). The second was pre-registered with
a refutation threshold of $+0.05$ and exceeds it at every seed, so a query may be read one
feed-forward early and may not be read from the network's input. That is where the method's
reach ends, and knowing the boundary is what the arm was run for.

\nSeeds{} seeds do not settle the shifted arm: its mean is inside the floor, but its seeds
disagree by \pairedSeedSpread{} --- more than the effect, and the widest disagreement here. Two
gave it a lead and the third reversed the sign. We report it and build nothing on it.

\paragraph{At this budget a lead is a disturbance, not a verdict.} These runs are at three per
cent of compute-optimal, where a difference in bits per byte is a difference in optimisation
trajectory rather than in reachable quality. A gap against the control measures how much the
change disturbed training, so the quantity is how far each arm moved, not which leads. The
\countA{} arms of family A move at most \maxDevA{} --- an order of magnitude inside the
\seedBand{} that \nSeeds{} seeds resolve, and smaller than an arm's own spread across seeds,
\seedSpread{}.

That is not a way of turning a null into a result: the same measurement resolves exchanging the
sub-layers at \rangeSwap{} times any read point's movement, and projecting the query from the
network's input at \rangeSingle{}. The instrument has range and family A produces no signal in
it. The flatness is not exact throughout --- the read points separate above the seed band from
roughly step 1{,}000 to 5{,}000 by at most half again, and the separation is gone by the last
third, so a campaign stopped at its widest point would have reported an effect a longer one does
not.

\paragraph{The families are not compared across.} Every family A arm sits below Standard, by
\gapSW{} to \gapKE{}. We list it and claim nothing from it, for three measured reasons: the two
stacks are the same alternating chain differing only in the half-layers at their ends, and the
shifted arm --- family B by construction, family A in its interior --- lands \shiftShare{} times
short of the gap; the tied embedding that is also the readout is frozen at its initialisation,
so one ordering hands that representation to a feed-forward first and the other to attention;
and the gap is a difference in convergence rate, \swapEarly{} at the first evaluation, widening
to \swapPeak{}, narrowing monotonically to \swapEffect{} by the end without having stopped. An
arm with the embedding unfrozen in both orderings would settle it; that changes a control this
campaign fixed. Nothing claimed here rests on the gap.

\paragraph{Parameters and arithmetic.} Every arm in both tables carries \armParams{} parameters
and identical forward arithmetic, measured rather than assumed, so no comparison here needs a
budget argument and none is offered. The parallel arrangement \citep{wang2021gptj,
chowdhery2023palm} is listed in family A because it shares the defining property, with the
caveat of Section~\ref{sec:algebra} that it is a different graph.

\paragraph{What \nSeeds{} seeds can and cannot say.} Every seed is printed. The tables are read
in paired differences, where the seed's effect on initialisation and data order is common to an
arm and its control and cancels; read that way the seeds scatter with a pooled standard
deviation of \pairedSd{}, putting a 95\% band of \seedBand{} on a mean of \nSeeds{}. That
measured resolution agrees with the floor of \runFloor{} fixed before these runs, and is what
makes family A a measurement rather than an absence: its arms sit within \maxDevA{} of their
control, an order of magnitude inside the band, while the two effects this paper rests on sit
\rangeSwap{} and \rangeSingle{} times outside it.

\section{Where a query may be read from, as algebra}
\label{sec:algebra}

Why the query, and why a read point rather than a different graph, are settled here rather than
assumed. Write the stream's two updates as operators, each carrying its own normalisation: $A$ for
attention, $M$ for the feed-forward. An ordinary block composes them; a block reading the query
earlier splits the attention's arguments, scoring from $x$ and loading values from $z$:
\begin{equation*}
S = (I + M) \circ (I + A), \qquad
Q = (I + A_{x;z}) \circ (I + M), \qquad z = x + M(x).
\end{equation*}
Over a stack the two are conjugate,
\begin{equation*}
Q_L = \big[(I+A)\circ(I+M)\big]^L = (I+M)^{-1} \circ S_L \circ (I+M),
\end{equation*}
so a stack of such blocks is the ordinary stack with its layer boundaries drawn half a sub-layer
over: the interiors agree exactly and only the two ends differ. We verify this constructively,
building the shifted stack as $L-1$ fused blocks between two half blocks and checking it is
bit-identical to the ordinary one.

The conjugation leaves each of attention's three arguments free to be read before the
feed-forward or after it, and the device selects among them: the sweep begins when the query
arrives, so $q$ from the earlier point buys the overlap, and $q$ alone. Where the key is read
changes nothing --- the device works over cached keys and the current one is a write nothing
waits on, holding to \keyLateRel{} relative for a block whose key is read late --- and a query
read from the later point is the ordinary block with its sub-layers exchanged, which gets no
overlap. Reading $q$ early is necessary and sufficient, and all a deployment changes. The
arrangement that sums rather than composes, $P = I + A + M$
\citep{wang2021gptj,chowdhery2023palm}, buys the same overlap by changing the graph instead;
a forward-pass test separates the three families, and that change is left to
separate work.

\section{Related work}
\label{sec:related}

Disaggregated serving splits a decode across machines by phase, prefill from decode
\citep{zhong2024distserve,patel2024splitwise}, or by tensor kind, the cache sweep from the static
weights \citep{lamina2024,adrenaline2025,instinfer2024}. A production line carries the second
split to its conclusion and disaggregates attention from the feed-forward outright
\citep{zhu2025megascale,stepfun2025step3} --- the split of Section~\ref{sec:twogpu} ---
overlapping the two sides across requests by micro-batching. Within one sequence the two still
alternate, and that is the dependency this paper removes; the two overlaps are orthogonal and
compose. Closest in framing, \citet{ma2026movequery} route the query itself between instances,
as DistAttention does for dense attention, shipping a query row with its softmax statistics
\citep{lin2024infinitellm}. What those inherit, and this
paper removes, is the block's own ordering. A separate line shrinks what the sweep must read
rather than when it may start --- one key-value head \citep{shazeer2019mqa} or a few
\citep{ainslie2023gqa}, or a low-rank code in place of the stored pair \citep{deepseekv2}; those
change the cache and compose with this, which changes neither cache nor kernel. The parallel
block \citep{wang2021gptj,chowdhery2023palm} buys the same overlap by changing the graph, and is
treated in Section~\ref{sec:algebra}.

\section{What was measured, and on what}
\label{sec:setup}

Each of the \nArms{} arms is \qsixName{} trained from scratch at \nSeeds{} seeds, every layer
converted, on \armTokens{} tokens of WikiText-103 over three epochs: \armSteps{} optimiser steps
at a peak learning rate of \armLR{} under a cosine schedule, sequence length \seqLen{},
\tokPerStep{} tokens per step, AdamW, compiled. The tied embedding, which is also the readout,
is frozen at initialisation so the comparison is between decoder stacks rather than
vocabularies. Two settings are pinned rather than left to a default, each because it would
otherwise dominate what the study resolves: normalisation weights are held in float32, since in
pure bfloat16 they begin at exactly $1.0$ where the spacing is $0.0078$ and the optimiser's step
is three orders of magnitude smaller, so they never move; and the attention backend is pinned to
flash and recorded per run: the alternative kernels disagree with it by up to \backendGap{} on
this model's attention shapes, which does not reach the loss here but is not a promise the
backend heuristic makes. Each run takes
\armHours{}~hours on one accelerator; the \nRuns{} runs here are \gpuHoursTotal{}~GPU-hours.

\paragraph{Which split these numbers are.} Every figure here is validation, and both the
checkpoint and the learning rate were selected on it. We therefore read the same runs on a
held-out test split that nothing selected on: against its own family's control, no gap moves
between splits by more than \splitMoveMax{} across \splitArms{} arm-seeds, and none in family A
by more than \splitMoveA{} --- both inside \runFloor{}. Selection did not manufacture the
flatness, and the refutation in the second table is larger on the split it was not selected on.

The validation split also shares windows with training. Each run scores a masked copy with those
windows removed: the effect is \leakEffect{}, varying by \leakSpread{} across \leakRuns{} runs.
A constant shift an order of magnitude below \runFloor{} in its variation cancels in every
comparison made here.

The protocol is verified against fused attention to \splitRel{} relative, on synthetic tensors
and end to end on a trained checkpoint, at the shapes and layers the release names.

\section{Conclusion}
\label{sec:conclusion}

The cache-side device of a disaggregated decoder needs the query and nothing else. That is an
identity, it runs on stock kernels, and what it returns is a mergeable aggregate, so a cache may
be split across any number of devices and merged in any order.

An existing checkpoint can be moved to this read point by fine-tuning $W_q$ alone, \convFrac{}\%
of the model, at a residual cost bounded by \convD{}~bits per byte at \convSteps{} steps
(Section~\ref{sec:convert}).

\paragraph{Reproducibility.} Every number here is generated from run records by a script in the
repository; none is typed into the source, and an unmeasured number raises a build error rather
than appearing as a placeholder. The block definitions, trainer, protocol tests and analysis are
included with the exact command for each arm. The pre-registrations are under
\texttt{protocols/}, one file per gate, each stating its verdict thresholds and the author's
prior before the arm ran; \texttt{gate\_SQ.md} is the one the $+0.05$ refutation threshold comes
from. The table each campaign was launched from is under \texttt{run/experiments/}, one row per
run.

\section{Limitations}
\label{sec:limits}

\begin{center}
\small
\begin{tabular}{p{0.44\textwidth}p{0.46\textwidth}}
\toprule
limitation & what would settle it \\
\midrule
One model family, one scale, one corpus, \nSeeds{} seeds, at three per cent of compute-optimal ---
where a difference in convergence and one in reachable quality are hardest to separate, and
every separation is still contracting at the last evaluation. &
More scales and corpora. What \nSeeds{} seeds do resolve is measured, not assumed: \seedBand{}.
\\
The arrangement is verified, not built. Correctness is an identity, checked end to end on a
trained checkpoint with the freshness condition stated in full; no speedup is measured. &
A deployment. Whether it pays is a ratio between an interconnect's latency and a feed-forward's
duration --- a property of the machine, not of this decomposition. \\
The family gap, \swapEffect{}, is reported and not explained (Section~\ref{sec:cost}). &
Both orderings with the embedding unfrozen, which changes a control this campaign fixed. \\
The conversion result is one budget on one corpus, and equal steps is not equal difficulty:
the converted arm spends part of its budget undoing the conversion, so \convD{} bounds the
residual cost from above rather than measuring it. &
The same arms at a larger budget, and on a second corpus. \\
\bottomrule
\end{tabular}
\end{center}

%
\renewcommand{\UrlBreaks}{\do\/\do\-\do\.\do\:\do\_\do\?\do\=\do\&}

\let\oldthebibliography\thebibliography
\renewcommand{\thebibliography}[1]{\oldthebibliography{#1}\raggedright}
\sloppy
\bibliographystyle{plainnat}
\bibliography{refs}

\ifdefined\LONGBUILD
\appendix

\let\qfirstoldsection\section
\renewcommand{\section}{\clearpage\qfirstoldsection}

\section{the combine, and what the disaggregated stage costs}
\label{app:combine}

\noindent The body states the join as a lerp weighted by a sigmoid, and states the
stage's cost in one line. This part carries both: where that form comes from, why the
logarithm is not a convenience, what the stage puts on the wire, and how wide the window
it opens really is.

\paragraph{Background: what the combine is.} Combining two partial softmax states \emph{is} a
sigmoid of the difference of their log partition functions, and the form says what the step means. For a fixed query the whole history
behaves as one pseudo-token of score $\mathrm{lse}_{<j}$ and value $o_{<j}$, so appending a token
is a two-way softmax --- a logistic. Thus $\mathrm{lse}_{<j} - s_{jj}$ is the log-odds that the
step attends to what it already has rather than to what just arrived. Decoding is a fold over
that pair,
\begin{equation*}
o_{<j+1} = \mathrm{lerp}\big(v_j,\, o_{<j},\, \sigma(\mathrm{lse}_{<j} - s_{jj})\big),
\qquad
\mathrm{lse}_{<j+1} = \mathrm{logaddexp}(\mathrm{lse}_{<j},\, s_{jj}),
\end{equation*}
which is the sense in which the device's two return values are sufficient: they are the state
of the fold.

\paragraph{Why the logarithm is not a convenience.}
Why it matters: in the undisguised coordinates a shard of \shardPositions{} positions at a score
spread of \shardSpread{} already carries $Z$ about \shardZ{} and $\|W\|$ about \shardW{}, an
order of magnitude below what single precision holds, so a wider spread or a longer shard cannot
be exchanged at all. Logged, $O$ is a convex combination and $\log Z$ a number near
\shardLogZ{}.
\paragraph{What it costs.} Statelessness puts the request's state on the wire and the layer loop
on the network. Per stage the cache side sends \stageOut{}~KB and receives \stageBack{}~KB ---
\tokenWire{}~KB per token over \nLayers{} round trips at \qsixName{}, \tokenWireBig{}~KB at
\qfourName{}. The bytes are immaterial; the serial round trips are the cost. What pays for them
is that the sweep begins the moment the request is dispatched, so a round trip is hidden whenever
it is shorter than a sweep and the margin grows with context. Across a wide-area link it does
not, and that is where the arrangement stops.

\paragraph{The window is one feed-forward wide.} A sweep runs concurrently with one feed-forward
and no more. Reading every layer's query from the network's input would widen that window to the
whole stack, since every query would exist at the step boundary --- but it costs
\singleQ{}~bits per byte at every seed (Section~\ref{sec:cost}). The preemptible pool is what
makes the limit survivable: a request away from the pool is not a pool standing idle.

\paragraph{How small the change is.}
Letting a block hand over its query early is a smaller change than it appears: the feed-forward
moves in front of the attention and the query keeps reading the block's input, leaving every
component, shape, parameter count, forward arithmetic and serial depth where they were. Over
\nSeeds{} seeds at \qsixName{}, no read point in that family separates from any other by more
than \maxDevA{}~bits per byte, against a resolution of \seedBand{}, while the same runs resolve
a sub-layer exchange \rangeSwap{} times as large and a shared query source \rangeSingle{} times
as large. The reading is a magnitude, not a ranking: at this budget an arm leading by a wide
margin would show the change perturbs training, not that it is good. The window is one
feed-forward wide and no wider --- a query projected from the network's input moves \singleQ{}.

\paragraph{What one card cannot answer.}
Whether the arrangement pays is a ratio between an interconnect's latency and a feed-forward's
duration, which one card cannot answer and building it would. The code is released for that.

\begin{figure}[t]
\centering
\includegraphics[width=0.80\textwidth]{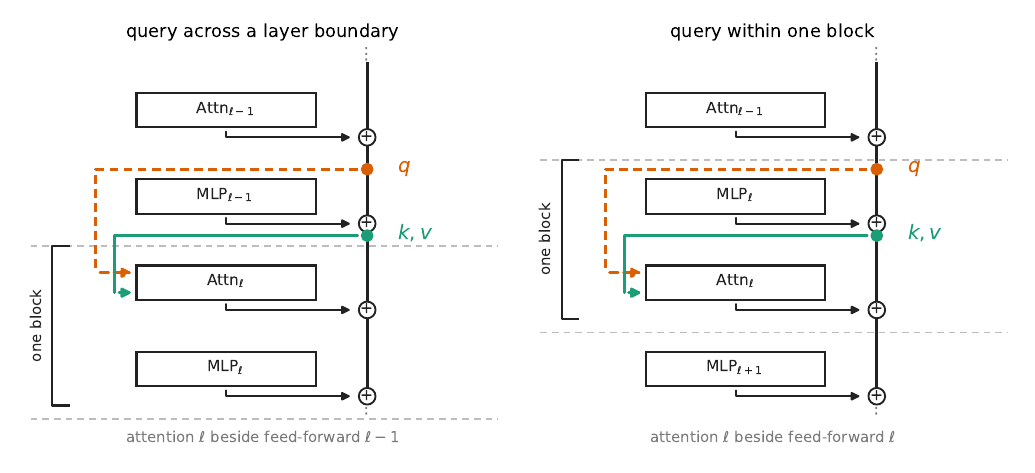}
\caption{The same wiring, bracketed two ways. Left: the query is tapped at the entrance of
feed-forward $\ell-1$ and crosses into attention $\ell$. Right: the feed-forward comes first and
the identical tap falls inside the block. The wires are identical; the bracket differs, and with
it which pretrained feed-forward sits beside which pretrained attention, plus a half block at
each end. Neither panel reads a staler query --- one feed-forward separates the query's tap from
the key's and value's in both --- and assuming otherwise attributes a pairing effect to
staleness.}
\label{fig:wiring}
\end{figure}
\section{the conversion and the depth sweep in detail}
\label{app:convert}

\paragraph{The second scale.}
At \qfourName{} the same procedure leaves \convDBig{}, and the untrained displacement is
\convUntrainedBig{} against \convUntrained{}. The share of the model $W_q$ occupies is
essentially unchanged between the two (\convFrac{}\% and \convFracBig{}\%), so the improvement
is not a budget effect; what produces it is not established here, and two scales are two points.

\paragraph{How the four depths were run.}
The same conversion and the same repair were run at four depths to establish that. Depths are in
HALF LAYERS: a layer is an attention and a feed-forward, and a query read one layer index back
moves half of one, so every reachable depth is odd. Layers \depthHeld{} are held out of the
conversion at every depth --- and layer 0 is exempt, as before --- leaving \depthLayers{}
converted in all four arms --- a layer with
fewer layers beneath it than the depth asks for cannot read that far and clamps to the bottom of
the stack, and without the hold-out the deeper arms would convert the same layers on paper while
some of them read somewhere else.

The shadow column is geometry, not measurement: consecutive softmax-attention layers in a
three-linear-one-attention pattern are four layers apart, so a depth of $h$ half-layers opens
$h/8$ of that interval, and $7$ half-layers --- the deepest odd depth inside it --- spans all
but the last eighth. The two costs beside it move in
opposite directions because a deeper arm starts further from its reference: it has more to remove,
removes a larger share of it, and still ends further away.

On a hybrid checkpoint the depths cost less to begin with. \hybridName{}, \hybridLayers{} layers
of which \hybridFull{} are softmax attention, is displaced \hybridOne{} bits per byte at one
half-layer and \hybridSeven{} at seven, against a baseline of \hybridBase{} --- against
\depthFreeSeven{} for the dense model at the same depth. That is forward-only: the checkpoint is
rewired and evaluated, with nothing trained.
\section{the design is forced, and each step is checked}
\label{app:forced}

\paragraph{The design is forced, and each step is checked.} Collected, the case that this is the
read point to move rests on four claims, each with the theory that predicts it and the
measurement that tests it:

\begin{center}
\small
\begin{tabular}{p{0.38\textwidth}p{0.50\textwidth}}
\toprule
claim & how it is established \\
\midrule
The shifted stack is the ordinary stack re-bracketed, not a different model. &
The conjugation $Q_L = (I+M)^{-1} \circ S_L \circ (I+M)$, verified constructively
bit-identical. \\
Only the query has to move early. &
The sweep runs over cached positions and the current key is a write nothing waits on, at
\keyLateRel{} relative for a block reading its key late. Moving the key or value instead buys
no overlap. \\
Moving it is free at this budget. &
\countA{} arms within \maxDevA{} of their control against a resolution of \seedBand{}, while the
same runs resolve effects \rangeSwap{} and \rangeSingle{} times larger
(Section~\ref{sec:cost}). \\
The freedom ends one feed-forward out. &
A query projected from the network's input costs \singleQ{}, exceeding a pre-registered
refutation threshold of $+0.05$ at every seed (Section~\ref{sec:cost}). \\
A checkpoint already in service reaches it far more cheaply than it reaches the alternative. &
Converted the same way and repaired by the weights whose input \emph{it} moves, the parallel
block needs \convParFracBig{}\% of \qfourName{} retrained against \convFracBig{}\% for the
query, is displaced \convParUntrainedBig{} against \convUntrainedBig{}, and retains
\convParDBig{} against \convDBig{} (Section~\ref{sec:convert}). \\
\bottomrule
\end{tabular}
\end{center}

The last three are what separates this from an arrangement chosen for convenience: the campaign
was run to find the boundary as much as the freedom, and it found one. The final row is about
reaching the two arrangements from a trained checkpoint and not about the arrangements
themselves --- the parallel block is one of the family A arms above that this campaign cannot
distinguish from any other, and it trains from scratch without difficulty.

\section{the wiring as a schedule, worked}
\label{app:schedule}

\noindent The algebra says which tensor is read where; what that does to a running server is the
next question. One instance, small enough to check by hand: \afdRequests{} requests over
\afdLayers{} layers, arriving in three groups one time unit apart, each with its own attention
time. The feed-forward is a pool per layer --- one unit per departure, one at a time, and it will
not depart with fewer than \afdMinBatch{} waiting. Communication is free.

Group G1 arrives at $t=0$ carrying R1(4), R2(3), R3(2), R4(1); G2 at $t=1$ with R5(3),
R6(2), R7(2), R8(1); G3 at $t=2$ with R9(2), R10(2), R11(1), R12(1), the bracketed
number being that request's attention time at every layer.

\paragraph{What the host may issue, and when.} Layer $\ell$'s query is projected from
$h_{\ell-1}$, the residual after layer $\ell-1$'s attention and before its feed-forward; the key
and value still come from $x_\ell$, which is after it. So the moment $h_{\ell-1}$ exists, two
things are issuable at once --- the sweep of layer $\ell$, which needs only $q_\ell$ and the cached
keys and values, and the feed-forward of layer $\ell-1$, which needs only $h_{\ell-1}$. Layer
$\ell$ is ready when both return, the second because $x_\ell = h_{\ell-1} +
\mathrm{MLP}_{\ell-1}$ is where its own key and value come from:
\begin{align*}
e_1 &= \text{arrival} + a, &&\text{layer 1's query reads the embedding; nothing to pair with} \\
e_\ell &= \max\big(e_{\ell-1} + a,\; \mathrm{ffn}_{\ell-1}\big), &&\ell = 2 \ldots \afdLayers{} \\
\text{finish} &= \mathrm{ffn}_{\afdLayers{}}, &&\text{the last feed-forward has no sweep left to pair with.}
\end{align*}
In a composed block that edge sits one half-layer later: the feed-forward of layer $\ell$ reads
$h_\ell$, so it cannot be issued until layer $\ell$'s attention has finished, and layer $\ell+1$
cannot begin until it returns.

\begin{figure}[!ht]
\centering
\includegraphics[width=\textwidth]{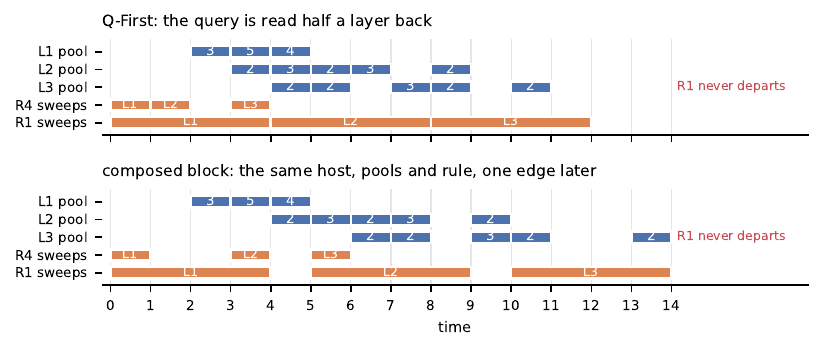}
\caption{One simulation, both wirings. Blue: each layer's pool, labelled with how many
requests are aboard; orange: the sweeps of the fastest request and the slowest. Read the bottom
row. Under Q-First the slowest request's sweeps abut --- $[0,4]$, $[4,8]$, $[8,12]$ --- because
its next query exists before the previous feed-forward runs; under the composed block a
feed-forward separates them at every boundary. The departures are the same batches in the same
order in both panels, only shifted.}
\label{fig:gantt}
\end{figure}

\paragraph{What the picture shows, and what it does not.} The last request to finish does so at
$t=\afdLastQF{}$ against $t=\afdLastComposed{}$, but that number is a property of these twelve
service times and is not a speedup this paper claims. What generalises is the shape: the sweep chain is
contiguous under one wiring and interrupted under the other by exactly one feed-forward at every
layer boundary. The pool itself is untouched --- same batches, same departures --- which is the
practical content of the claim that nothing but the query moves.

\paragraph{One hazard the instance exposes.} The departure rule as stated has no timeout, and
\afdStranded{} request never leaves: R1 is the slowest and the last, so when its final
feed-forward request arrives there is nothing left to ride with. It happens identically under both
wirings, so it belongs to the rule and not to the block, and to a finite example draining out
rather than to a server in steady state. An implementation needs a maximum wait, and that wait
enters the latency.

\section{two checks before trusting the arithmetic}
\label{app:decide}

\noindent The arrangement that sums rather than composes,
$P = I + A + M$ \citep{wang2021gptj,chowdhery2023palm}, delivers the same overlap and is not the
same kind of change. $S$ and $Q$ differ in where three arguments are read; $P$ differs in the
graph. Its branches never meet inside the block, so an $L$-layer stack has $L$ sub-layers on its
longest dependency chain where $S$ and $Q$ have $2L$.

Which of the three a block belongs to is decidable. For a block $B$ taking its two sub-layers
scaled, the mixed second difference
\begin{equation*}
\Delta = \big[B(\alpha A, \mu M) - B(\alpha A, M)\big] - \big[B(A, \mu M) - B(A, M)\big]
\end{equation*}
vanishes identically exactly when the two branches never meet. Measured on this model, it is
zero for $P$ and nonzero for $S$ and for every $Q$.

We raise the distinction rather than pursue it because only one of the two changes is free.
Halving a stack's serial depth changes what the model can compose; moving a read point does not.
The serial-depth change is left to separate work.

\paragraph{Three numerical caveats.}

\noindent The claims in Section~\ref{sec:protocol} rest on these holding.

\begin{center}
\small
\begin{tabular}{p{0.31\textwidth}p{0.59\textwidth}}
\toprule
where & why \\
\midrule
the running log partition &
$\log(e^a+e^b)$ overflows at $a=b=100$, an ordinary magnitude at long context; a stabilised
$\mathrm{logaddexp}$ is required. \\
indexing the fused kernel's log-sum-exp &
it is padded along the sequence axis with non-finite entries, so it must be read at the query's
position, not reduced over it. A reduction poisons the fold with an infinity from no score. \\
the sign of $s_{jj}-\mathrm{lse}_{<j}$ &
$\mathrm{lse}_{<j}$ spans only the cache, so $s_{jj}$ may exceed it --- on a trained checkpoint,
in \shortCacheOver{} of head-positions at cache length one, by up to \shortCacheNats{}~nats.
Short caches are the first tokens of every sequence, so a one-directional form handles none. \\
\bottomrule
\end{tabular}
\end{center}

\section{tuning the split}

\begin{figure}[t]
\centering
\includegraphics[width=0.76\textwidth]{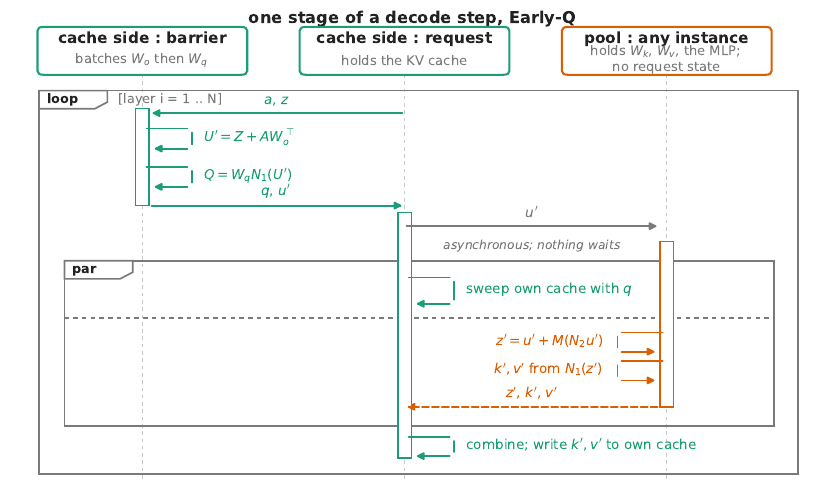}
\caption{One stage of an Early-Q decode, as an interaction. Each lifeline's subtitle says what
that participant holds. The barrier and the request are the same device, drawn apart because the
projections are batched across whichever requests are ready while the sweeps run free and rejoin
later; the \texttt{par} fragment is the dispatch the sweep does not wait for.}
\label{fig:uml}
\end{figure}

None of the following bears on a claim in the body. It is what a deployment has to
decide once it has accepted the contract of Section~\ref{sec:master}, collected here so
that the body can stay on the one thing it measures.

\paragraph{The cache side must batch.} Served alone,
the two projections it owns are matrix-vector products --- one operation per byte read --- and
cost \pairGemv{}~ms per token at \qsixName{} and \pairGemvBig{}~ms at the \qfourName{} shape,
against \pairBatched{}~ms and \pairBatchedBig{}~ms over a batch of eight. Batching is therefore a
precondition of this placement and not an optimisation on top of it; a deployment unwilling to
batch should pool the pair instead. One barrier per layer suffices, because $W_o$ closes stage
$i$ and $W_q$ opens stage $i+1$ with only a residual add and a normalisation between them
(Figure~\ref{fig:uml}). The barrier holds those two products and not the sweeps: contexts differ
by orders of magnitude, and a wave that locked the sweeps together would run at the pace of its
longest member.

\paragraph{When to pool the pair
instead.} Both projections may go to the pool, which then returns the projected query as a first
reply and the feed-forward's results as a second. The cache side holds no weights at all: a pure
sweep engine, replicating nothing across a cache tier and pinned to no model, so one fleet may
serve several. It costs two replies, twice the wire, a round trip before every sweep, and
\pairShare{}\% more arithmetic on the pool. Which is right is a question about concurrency, and
the two effects run opposite ways: the round trip is a stall another request fills, while the
pool's extra arithmetic scales with concurrency and never amortises. Below roughly
\breakEven{} co-resident requests at \qsixName{} (\breakEvenBig{} at \qfourName{}) pool the pair;
above it, keep it local.

\textbf{Keeping $W_o$ also unloads the resource that saturates.} The two sides do not saturate
alike. A weight read amortises over the requests sharing a device; a cache sweep does not,
because each request sweeps a cache of its own. So $W_o$ on the cache side costs its bandwidth
divided by the number of co-resident requests, and is paid for once past
\breakEven{} of them at \qsixName{} and \breakEvenBig{} at the \qfourName{} shape, even when the
cache side is an order of magnitude slower than the pool. What it buys does not amortise away:
\poolShare{}\% of a layer's arithmetic at \qsixName{} and \poolShareBig{}\% at \qfourName{},
removed from the pool, which is the shared resource and therefore the one that runs out first
under concurrency.

\textbf{It also deletes the watermark.} Section~\ref{sec:twogpu} has the compute side writing
into the cache side's memory, so it needs a per-layer write watermark to stop a prefetch reading
a cache one position short. A pure function returns its results instead: no shared mutable state
is left, and the check that guarded it has nothing to guard.

\section{what a query carries, and whether it compresses}

Nothing here is needed for the protocol or for the measurement in Section~\ref{sec:cost}, which
establishes what moving a read point costs by training the arms rather than by reasoning about
the projections. These are a second and independent line on the same question --- how much a
query's staleness can matter --- kept out of the body so that the body stays on the one claim it
measures. Every figure is forward-only, on released weights, with only the query touched.

\paragraph{A query is not a coordinate.} It was put to us that a query's job is to say WHICH ---
which token, which layer --- and not to carry what a token means. Rotary embedding supplies the
token coordinate whatever $W_q$ produces, so what remains testable is whether $W_q N(x)$ is close
to a per-layer constant. Replace each layer's query with its own mean over tokens, which
preserves the layer exactly and removes every trace of content:

\begin{center}
\small
\begin{tabular}{lrrr}
\toprule
model & baseline & query replaced by its layer mean & between-layer share of variance \\
\midrule
\qsixName{}  & \convRefBpb{} & \meanQBpb{} \;(\meanQCost{})    & \betweenLayer{}\% \\
\qfourName{} & ---           & \meanQBpbBig{} \;(\meanQCostBig{}) & \betweenLayerBig{}\% \\
\bottomrule
\end{tabular}
\end{center}

The coordinate is entirely intact and the model is destroyed, by an order of magnitude more than
the whole conversion costs. The large between-layer share does not rescue the proposal either:
a layer's query meets only that layer's keys, so a tag saying which layer this is has no reader
downstream.

\paragraph{What the displacement should be measured against.} The conversion moves a query by
about a third of its norm. That figure cannot separate two opposite situations, because a query
is a common component plus a token-dependent one and only the second is legible to a score. The
denominator that matters is the spread across tokens within a layer:

\begin{center}
\small
\begin{tabular}{lrrr}
\toprule
model & displacement / norm & displacement / spread & common share of the query \\
\midrule
\qsixName{}   & \dispNorm{}       & \dispSpread{}       & \commonShare{}\% \\
\qfourName{}  & \dispNormBig{}    & \dispSpreadBig{}    & \commonShareBig{}\% \\
\hybridName{} & \dispNormHybrid{} & \dispSpreadHybrid{} & \commonShareHybrid{}\% \\
\bottomrule
\end{tabular}
\end{center}

Both fall with scale and they fall at different rates, because the common share rises: the larger
model's query sits more in a part no score can distinguish. Neither series tracks the conversion
cost closely enough to be offered as its explanation --- that cost falls by a factor of four over
the same three checkpoints while the ratio falls by about a fifth --- and it is reported here as a
measurement rather than as a mechanism.

\paragraph{Tolerance to a query error, and what carries it.} Perturbing a query by the
displacement a conversion makes, in a random direction, costs a fraction of what the conversion
itself costs; the ratio is within one model and does not depend on where its baseline sits.

\begin{center}
\small
\begin{tabular}{lccr}
\toprule
model & latent KV & sparse routing & conversion / random \\
\midrule
\qsixName{}, \qfourName{}       & no  & no  & \Tol{}, \TolBig{} \\
\texttt{Qwen3-30B-A3B}          & no  & 128 experts & \TolMoe{} \\
\texttt{Youtu-LLM-2B}           & yes & no  & \TolMla{} \\
\texttt{Moonlight-16B-A3B}      & yes & 64 experts & \TolBoth{} \\
\bottomrule
\end{tabular}
\end{center}

The split follows the latent cache and not the routing: a mixture of 128 experts sits in the
dense band, and a dense model with a compressed cache does not. A ratio near one says the
conversion is doing nothing a random perturbation of the same size does not --- those
checkpoints are intolerant of query error however it arrives, and a deployment converting one
should expect the training-free cost to be large before any repair.

\section{Why an early query is the cheap thing to want}
\label{sec:softmax}

A protocol that starts on the query is worth having only if a query is cheap to produce early.
One half of that is a bound, and needs no data. Writing $o = \sum_j A_j v_j$ with
$A = \mathrm{softmax}(s)$,
\begin{equation*}
\frac{\partial o}{\partial v_j} = A_j, \quad \textstyle\sum_j A_j = 1,
\qquad\qquad
\frac{\partial o}{\partial s_j} = A_j\,(v_j - o).
\end{equation*}
The value path has total gain exactly one. The query path is second order and vanishes at both
ends: sharp attention drives $A_j$ to zero away from its peak, and diffuse attention drives
$(v_j - o)$ to zero. Softmax is also invariant to a constant added to every score, so whatever
part of a query's change is uniform across keys does nothing at all.

That bounds the sensitivity but says nothing about the size of the perturbation, which is a
question about trained weights and is taken up in Section~\ref{sec:geometry}, after the cost of
the change itself has been measured.

\paragraph{The bound predicts something checkable, and it holds.} If a stale query hurt because
the information it is missing has to be made up from somewhere, the damage would fall as context
accumulates: at the first position there is nothing to make it up from, and by the thousandth
there are a thousand completed blocks behind it. Measured on \qsixName{}, the cost as a share of
that position's own loss does not decay with context: it rises early and then holds flat:

\begin{center}
\small
\begin{tabular}{lrrrr}
\toprule
positions & \posEarlyLo{}\,to\,6 & 76\,to\,181 & 181\,to\,431 & 431\,to\,\posLateHi{} \\
\midrule
cost, as a share of the loss at that position & \posEarly{}\% & 7.8\% & 7.5\% & \posLate{}\% \\
\bottomrule
\end{tabular}
\end{center}

Position zero is excluded: with one entry in the cache, attention puts all its mass there
whatever the query says, so the cost is zero by construction rather than by measurement. What the
flat tail rules out is the reading that earlier positions' values compensate a stale query --- if
they did, the share would fall with context, and over the measured range --- positions
\posEarlyLo{} to \posLateHi{} --- it does not. The cost of an early query is set by how far back it is read and not by how much has
been read before it.

\section{What training puts in the projections}
\label{sec:geometry}

What follows leaves the deployment argument behind and asks what a trained model puts in a
query --- which the protocol does not need answered, and which is the natural question of a paper
that has just moved a query's read point. Section~\ref{sec:softmax} bounded the sensitivity; what
remains is the size of the perturbation a feed-forward makes \citep{geva2021ffn}, measured on the
official
\qsixName{} release --- the only fully trained model here and the one a deployment loads ---
against a randomly initialised model of the same shape. The null matters more than the
measurement: several of these statistics have values at initialisation that a reader would
take for structure.

\paragraph{All three projections collapse; how far the value collapses depends on the budget.}
We report the stable rank, $\|W\|_F^2 / \|W\|_2^2$, rather than a count of singular values
above a threshold, because a threshold is a tunable knob:
\begin{center}
\begin{tabular}{lccc}
\toprule
 & $W_q$ & $W_k$ & $W_v$ \\
\midrule
random initialisation      & \srQrand{} & \srKrand{} & \srVrand{} \\
official release           & \srQ{}     & \srK{}     & \srV{}     \\
a \armTokens{}-token model of the same shape & \srQours{} & \srKours{} & \srVours{} \\
\bottomrule
\end{tabular}
\end{center}
Out of $d = \dModel{}$, training takes all three a long way down. In the official release the
value keeps roughly three times what the query and key do; in a model of the same architecture
trained on \armTokens{} tokens --- three per cent of compute-optimal --- it does not, and its
stable rank sits beside theirs at \srVours{}.

We report the second row because the first alone would read as a claim about transformers in
general, whereas it is a claim about \emph{trained} transformers. The separation between the
value and the other two is something a model acquires with budget, not something the
architecture imposes, and a reader measuring this on a small model of their own would
otherwise find our number missing and conclude that their measurement was wrong.

\paragraph{The query binds to the key more than to the value.} We take each projection's
dominant subspace and ask how much of another projection's energy lies inside it, against the
chance level that a subspace of that dimension gives:
\begin{center}
\begin{tabular}{lcc}
\toprule
 & official release & \armTokens{}-token model \\
\midrule
$W_k$ inside $W_q$'s subspace & \kqRatio{} & \kqOurs{} \\
$W_v$ inside $W_q$'s subspace & \vqRatio{} & \vqOurs{} \\
\bottomrule
\end{tabular}
\end{center}
as ratios to chance, in every one of \nLayers{} layers and at both budgets. The two sides of a
score share geometry, as an inner product needs them to, and more than the value does. The
ordering survives the change of budget; the magnitudes do not, which is why both columns are
here.

This is worth stating because the family of head-sharing designs --- one key-value head for
many query heads \citep{shazeer2019mqa}, or a few \citep{ainslie2023gqa} --- treats the key
and the value as one object to be shared, whereas the geometry says they are learned as
different kinds of objects.

\paragraph{The low rank does not license a factorisation.}
\label{sec:truncation}

A stable rank of \srQ{} out of \dModel{} invites the obvious inference: factorise the query
and key projections through a rank-$r$ bottleneck and save the difference. Compressing the
attention's projections into a low-rank latent is a design in use \citep{deepseekv2}, so the
question is not whether it can be done but whether these matrices in this model permit it.

They do not. Truncating $W_q$ and $W_k$ to rank \truncRank{} on the official checkpoint costs
\truncCost{}~bits per byte; truncating $W_v$ to the same rank costs \truncCostV{}; and at rank
$512$, half the width, the query and key truncations still cost \truncHalf{}~bits per byte.
Stable rank is a concentration statistic and not a rank: the same matrices need \energyRank{}
directions to hold $99\%$ of their energy, so the spectrum is concentrated and the function is in
its tail.

The two are not in tension: a low rank says a generic perturbation reaches the query weakly,
and the truncations say the directions it does reach are ones the model cannot be deprived
of.

\fi

\end{document}